\documentstyle[aps,preprint,epsfig]{revtex}
\begin{document}
\draft{}
\preprint{LA PLATA TH-98/08}

\title{Cylindrically symmetric spinning Brans-Dicke spacetimes
with closed timelike curves}
\author{Luis A. Anchordoqui, Santiago E. Perez Bergliaffa  and Marta
L. Trobo}
\address{Departamento de F\'\i sica, Universidad Nacional de La Plata\\
 C.C. 67, (1900) La Plata\\
Argentina}
\author{Graciela S. Birman}
\address{Departamento de Matem\'atica, Fac. Cs. Exactas, UNCPBA\\
Pinto 399, (7000) Tandil\\
Argentina}
\maketitle
\begin{center}
\end{center}
\begin{abstract}
We present here three new solutions of Brans-Dicke theory for a
stationary geometry with cylindrical symmetry in the presence of
matter in rigid rotation with $T^\mu\,\!\!_\mu\neq 0$. All the solutions have eternal
closed timelike curves in some region of spacetime the size of which depends on $\omega$.
Moreover, two
of them do not go over a solution of general relativity in the
limit $\omega \rightarrow\infty$.
\end{abstract}

PACS number(s): 04.50+h, 04.20Jb

\vspace{1.5cm}

\newpage

General Relativity (GR) has survived many experimental tests during
its rather long existence. However, there remain some conceptual
difficulties (mostly related to Quantum Gravity) that have led to
the construction of several alternative theories (for a review, see
\cite{altt}). Brans-Dicke theory (BDT) \cite{BD} (along with its
generalization, namely the family of scalar-tensor theories
\cite{scaten}) seems to be the most promising of these because of
its excellent agreement with experiment and its intimate relationship with the
low energy limit of string theory \cite{strings}. To be more
precise, BDT is consistent with solar-system experiments if
the coupling parameter satisfies the inequality
$|\omega |>500$ \cite{will}. As a consequence, a great deal of
effort has been devoted to explore the whole space of solutions of BDT.
We will be concerned here with
solutions that display causal anomalies in the form of
closed timelike-curves (CTCs).
It has been known for a long time that CTCs are a
distinctive feature of geometries in which the light cones may be
tilt in an appropriate fashion.
The most widely known
geometries of these type are the so-called wormholes \cite{motho} which are exact
solutions of a given theory of gravitation with non-trivial topology (see \cite{visser} for
wormholes in GR, and \cite{bdwh} for the case of BDT). They
do not exhibit CTCs {\em ab initio}; some (rather artificial)
procedures to generate them are explained in \cite{wh}.
On the other hand, the desired tilting is naturally achieved in rotating
space-times such as the spinning black hole of Kerr \cite{kerr},
the infinite dust cylinder of van Stockum
\cite{vanS}, the universe of G\"odel \cite{goedel}, and their generalizations
\cite{ctc}. In this paper we present solutions to BDT
for a cylindrically symmetric spacetime with rigidly rotating matter modeled
by a stress-energy tensor with nonzero tangential stresses.
We shall see that the solutions have eternal closed timelike
curves in a certain region of spacetime, the boundary of which
depends on the value of $\omega$. We also
analyze the singularity structure of the solutions. Finally, we
study the $\omega\rightarrow\infty$ limit, and we find that two of
our solutions do not go over solutions of GR.

\hfill

If for simplicity units are chosen which make the velocity of light
equal to unity, the field equations of BDT are \footnote{In what follows,
Greek indices run from 0 to 3 while Latin indices from 1 to 3.
Semicolons denote covariant derivative with respect to the metric
$g_{\mu\nu}$, primes derivatives with respect to $r$, and the commas
mean partial derivatives with respect to the coordinate $x^\mu$.
As usual, $g \equiv$ det $g_{\mu\nu}$ and $\delta^\nu\,\!\!_\mu$ is the
Kronecker delta. Sub-indices with hat refer to the proper reference
frame, in which $g_{\hat{\mu}\hat{\nu}} \equiv \eta_{\mu\nu}$,
being $\eta_{\mu\nu}$ the metric of Minkowski spacetime.}
\cite{BD}
\begin{equation}
G_{\mu\nu}  =
\frac{8\pi}{\phi} T_{\mu\nu} + \frac{\omega}{\phi^2}
(\phi_{;\mu} \phi_{;\nu} - {1\over2} g_{\mu\nu} \phi^{;\alpha}
\phi_{;\alpha}) + \frac{1}{\phi} (\phi_{;\mu\nu} - g_{\mu\nu} \,\phi^{;\alpha}
\,\!\!_{;\alpha}),
\end{equation}
\begin{equation}
\Box \phi\equiv \phi^{;\alpha} \,\!\!_{;\alpha}= \frac{8\, \pi\, T}{3\,+\,2\,\omega}
\end{equation}
where $G_{\mu\nu}$ is the Einstein tensor,
$T_{\mu\nu}$ is the energy-momentum tensor of
matter and all nongravitational fields, and $T=T^\mu\,\!\!_\mu$.
In a stationary spacetime  the metric and
the scalar field can be chosen in such a way that $ g_{\mu\nu,0} =
\phi_{,0}$ = 0, and $g_{0i} = 0$. Further simplification is possible
if we assume that the geometry has cylindrical symmetry. The line element can be
written in this case as follows:
\begin{equation}
ds^2 = dt^2 - e^{2\lambda(r)} (dr^2 + dz^2) - l(r) d\theta^2 + 2 m(r)
d\theta dt.
\end{equation}
We shall take for the stress-energy tensor the expression,
\begin{equation}
T_{\mu\nu} = \rho u_\mu u_\nu + \Pi_{\mu\nu}
\end{equation}
with $\rho$ the density of energy, and
$u^\mu = \delta^\mu\,\!\!_0$ the four velocity in the comoving system.
$\Pi_{\mu\nu}$ is
the anisotropic pressure tensor, which is symmetric, trace-free, and
orthogonal to the comoving observer.
We will restrict ourselves here to the case in which
$\Pi_{\mu\nu} =$ diag$(0,-\alpha,\alpha,0)$ \footnote{Let us remark that a solution for
the same geometry in the case of dust in BDT was found by Bandyopadyhay \cite{niki},
but it exhibits the undesirable feature of a singularity at a finite radial proper
distance from the origin.}.

With the foregoing assumptions,
the nonvanishing field equations are
\begin{mathletters}
\begin{equation}
 \frac{d}{dr} \left( \frac{m m'}{2D} \right) = \frac{4
\pi }{\phi}\sqrt{-g} \rho  - \frac{m m'}{2 D} \frac{\phi'}{\phi} +
{1\over2}  \frac{ \Box
\phi}{\phi} \,\sqrt{-g}
\label{1}
\end{equation}
\begin{equation}
 \frac{d}{dr} \left( \frac{l' + m m'}{2D} \right) =
- \frac{4 \pi }{\phi} \sqrt{-g} \rho  - \frac{m m'
+ l'}{2D} \frac{\phi'}{\phi} + {1\over2}  \frac{
\Box\phi}{\phi}\,\sqrt{-g}
\label{2}
\end{equation}
\begin{equation}
\frac{d}{dr} \left(\frac{m'}{2D} \right) = - \frac{m'}{2D} \frac{\phi'}{\phi}
\label{3}
\end{equation}
\begin{equation}
 \frac{d}{dr} \left( \frac{m l' - m' l}{2D} \right) = -
\frac{ 8 \pi}{\phi} \sqrt{-g} m \rho  +  \frac{m' l - m l'}{2D}
\frac{\phi'}{\phi}
\label{4}
\end{equation}
\begin{equation}
-D \lambda'' + \frac{m'^2}{2D} + D'\lambda' - D''
=  \frac{4 \pi }{\phi} \sqrt{-g}(\rho - 2 \alpha) + \omega D \frac{\phi'^2}
{\phi^2} + D  \frac{\phi''}{\phi} - D \lambda' \frac{\phi'}{\phi} -
{1\over2}  \frac{\Box \phi}{\phi}\,\sqrt{-g}
\label{5}
\end{equation}
\begin{equation}
-D \lambda'' - D' \lambda' =  \frac{4 \pi}{\phi}\sqrt{-g}
(\rho + 2 \alpha) + D \lambda'
\frac{\phi'}{\phi} - {1\over2}
\frac{\Box \phi}{\phi}\,\sqrt{-g}
\label{6}
\end{equation}
\begin{equation}
\Box \phi = \frac{8 \pi}{3+ 2 \omega} \rho
\label{7}
\end{equation}
\begin{equation}
D^2=l+m^2.
\label{8}
\end{equation}
\end{mathletters}
To solve this system, first note that Eq. (\ref{3})
readily gives the first integral
\begin{equation}
\frac{m'}{2D} = \frac{b}{\phi}
\label{12}
\end{equation}
where $b$ is an integration constant.
From Eqs. (\ref{1}) and (\ref{7}) we obtain at once that
\begin{equation}
\frac{2\,D\,b^2}{\phi} = (2 + \omega) \sqrt{-g}\: \Box \phi
\end{equation}
whereas from the definition of $\Box \phi$ one has
\begin{equation}
D  =  \exp \left\{- \int \frac{2\,b^2\,+(2+\omega) \phi \,\phi''}{(2+\omega)
\phi\,\phi'} dr \right\}
\label{11}
\end{equation}
Eq. (\ref{4}) in conjunction with Eqs.
(\ref{8}), (\ref{12}) and (\ref{11}) determine a differential
equation for the BD scalar. We shall propose
three different types of solutions. In the case of Type I solutions
we assume that the scalar field is given by a polynomical expression in
$r$; in Type II, by a constant times a trigonometric function, and in Type III we assume that
$\phi$ is given by a constant
times an hyperbolic function. After an appropriate fixing of the constants to recover GR
whenever it is possible, we obtain the expressions of $\phi (r)$, $l(r)$ and $m(r)$
listed in Table I, along with the range of values of $\omega$ in which
each solution is valid. Finally, adding Eqs. (\ref{5}) and (\ref{6}) we obtain an
equation for $\lambda$,
\begin{equation}
\lambda'' = \frac{\sqrt{-g} (\omega+1)}{D} \frac{\Box \phi}{\phi} -
\frac{\omega}{2} \frac{\phi'^2}{\phi^2} - {1\over2}
\frac{\phi''}{\phi} + \frac{m'^2}{4D^2} - \frac{D''}{2D}
\end{equation}
Integrating this we obtain the expressions listed in Table II.

In order to analyze the singularity structure of these spacetimes,
we next study the behaviour of certain scalars
built from the Riemann tensor. It is easier to compute this
tensor in the orthonormal frame. Let us define the differential forms
\begin{equation}
\Theta^0 = - m \, d\theta + dt \hspace{.7cm} \Theta^1 = e^{\lambda} dr
\hspace{.7cm} \Theta^2 = e^{\lambda} dz
\hspace{.7cm} \Theta^3 = D \, d\theta
\end{equation}
with the corresponding basis vectors
\begin{equation}
e^i\,\!\!_{\hat{0}} = \delta^i\,\!\!_0 \hspace{.7cm} e^i\,\!\!_{\hat{1}} =
e^{-\lambda}\, \delta^i\,\!\!_1 \hspace{.7cm} e^i\,\!\!_{\hat{2}} =
e^{-\lambda}\, \delta^i\,\!\!_2 \hspace{.7cm} e^i\,\!\!_{\hat{3}} =  D^{-1}\,
\delta^i\,\!\!_3\, + \,m D^{-1}\, \delta^i\,\!\!_0.
\end{equation}
The relationship between the Riemann tensor in the two frames
is the usual one:
$$R_{\hat{\mu}\hat{\nu}\hat{\eta}\hat{\delta}} \,\,= e^\alpha\,\!\!_{\hat{\mu}}\,\,
e^\beta\,\!\!_{\hat{\nu}}\,\, e^\epsilon\,\!\!_{\hat{\eta}}\,\,e^\psi\,\!\!_{\hat{\delta}}
\,\,R_{\alpha\beta\epsilon\psi}$$

In the following we shall see that the first two solutions of
Table I are unphysical. From the
expression of $R_{\hat{\mu}\hat{\nu}\hat{\eta}\hat{\delta}}$ for
the polynomial case, it can be seen that there is a singularity at
the origin unless $\Upsilon = 0$. This implies that $\alpha=0$, and
so the solution reduces to the dust case previously analyzed in
\cite{niki} \footnote{Actually, one still has to require
that spacetime be Euclidean near the origin. This fixes $C$ to zero.}.
It is worth recalling that in this case
there is a singularity which occurs at a finite proper distance
from the axis of symmetry, since $\int_0^\infty \exp(\lambda) dr$
is convergent. Solutions of Type II has instead naked singularities
for $r=n\pi/2$, with $n\in {\cal N}$. Consequently, from now on, we shall
concentrate only on Type III solutions (valid for $\omega\in (-\infty , -2)$)
for which the components of the Riemann tensor are well behaved throughout the
entire spacetime.

The result of the calculation of the kinematical quantities associated with the fluid
shows that the expansion scalar, the acceleration vector,
and the shear tensor are all zero, but there is a nonzero vorticity vector given
(in the case of Type III solutions)  by
\begin{equation}
\omega^\mu = \left( 0,0,-\sqrt{|\omega| /2-1}\:(\cosh
r)^{-3-2\omega}\;e^{(1+\omega /2)r^2},0\right)
\end{equation}
Let us turn now to the analysis of the relation between cause and effect
in our spacetime. First, it is necessary to remark that, although in the
metric of Type I the coordinate $\theta$ is an angular
coordinate,
this is not the case in Type III. However, after a trivial coordinate
transformation the line element may be written in the form
\begin{equation}
ds^2= dt^2 - e^{2 \Lambda(r)}\,(dr^2 + dz^2) - L(r) \, d\vartheta^2 + 2 M(r)\,
dt\,d\vartheta
\end{equation}
where
$$\Lambda(r)=(1 + \omega) \ln(\cosh r) - (2+\omega)\,r^2/4,$$
$$L(r) =F^2(r)[1 + 2 (2+\omega) \sinh^2 r]\,\,{\rm sech}^2 r,$$ and
$$M(r) = \sqrt{2|\omega |-4}\,\, F(r) \,\,{\rm tanh} r,$$
where $F(r)$ is any function that goes to 0 as $r^2$.
The coordinate $\vartheta$ is now an angular coordinate. Consequently, any curve
with constant $t$, $r$, and $z$ is closed. In particular, such closed curves are timelike
(but they are not geodesics)
if $r>r_{\rm crit}={\rm arcsinh} [1/(4+2\omega)]$.

In order to characterize in more detail the matter that generates the geometry,
we determine next whether it violates the weak energy condition \cite{hawpen}
\footnote{We stress that we are concerned here only with the matter sector of the
stress-energy tensor. Note however that one should take into account both the BD and
the matter sectors of the stress-energy tensor in order to relate violations of WEC
to the behaviour of geodesics.}
To do so, we calculate $\alpha (r)$ from Eq. (\ref{6})
\begin{equation}
4 \pi \alpha = -b \,\frac{\sqrt{2 (|\omega|-2)}}{4}
(\cosh r)^{2|\omega|-1} \exp \left\{-\frac{|\omega|-2}{2}\;r^2 \right\}
\end{equation}
Finally from (\ref{7}), $\rho$ comes out as
\begin{equation}
4 \pi\rho =\frac{b}{\sqrt{2(|\omega|-2)}} (2|\omega|-3)
(\cosh r)^{2|\omega|-3} \exp \left\{-\frac{|\omega|-2}{2}\;r^2 \right\}
\label{rho}
\end{equation}
It is straightforward to compute that $\rho > 0, \forall r$ if $b>0$.
Thus matter will satisfy WEC if and only if $\rho \pm \alpha > 0$ or
equivalently if $r>r_{\rm WEC}={\rm arcosh}
\left[\sqrt{\frac{2(2|\omega|-3)}{|\omega|-2}}\right]$.

From Fig. 1, we can see that the radius of the causal region decreases fast with
$\omega$, and that there will be CTCs both in the presence of ``exotic'' and
``non-exotic'' matter \cite{motho} for all values of $\omega$.
It can also be seen that the
radius of the region in which matter violates WEC tends rapidly to a
constant.

We close with some remarks regarding the limit $\omega\rightarrow\infty$.
The study of the limits of geometries depending on some parameter has been initiated by
Geroch \cite{geroch}, who pointed out that such limit may depend
on the coordinate system chosen to perform the calculations. More
recently, Paiva {\em et al} \cite{paiva} developed a
coordinate-free approach (based on the characterization of a given space-time by the
Cartan scalars \cite{cartan}) to study these limits in GR. Romero and Barros found later
\cite{rombar} some examples of BD solutions which do not reduce to
GR when $\omega\rightarrow\infty$. This motivated the application
of the above mentioned coordinate-free method to BDT by Paiva and
Romero \cite{paiva2}. The method involves a lengthy calculation (crucial to
make any statement about the limit $\omega\rightarrow \infty$) which will
be presented elsewhere. Note however that the BD scalar in solutions of Type II and
III goes to zero in the limit $\omega\rightarrow\infty$, and so the effective
gravitational constant $G_{\rm eff}=G_0/\phi$ diverges. Clearly our solutions
do not go over GR, and are
then counterexamples to the recent claim by Banerjee and Sen that
the condition $T\neq 0$ is a necessary and sufficient condition
for BD solutions to yield the corresponding solutions with the
same energy-momentum tensor \cite{sen}.\\

\acknowledgements Special thanks go to D. Torres for a critical reading of
the manuscript. L.A.A. would like to thank FOMEC for financial support. S.E.P.B. was
supported by UNLP. M.L.T. and G.S.B were supported by CONICET.

\begin{table}
\caption{The table shows different choices of the BD scalar $\phi$,
the corresponding metric coefficients, and the maximum
range of $\omega$ for which each solution exists.}
\begin{tabular}{cccc}
$\phi(r)$ & $l(r)$ & $m(r)$ & Domain of $\omega$ \\ \hline
$1 - r^2 b^2/(4+2\omega)$ & $(r^2- b^2r^4)\,[1-r^2b^2/(4+2\omega)]^{-2}
$& $br^2 [1-r^2b^2/(4+2\omega)]^{-1}$   & $(-\infty, -2)$\\
$(b\,\sqrt{2}/\sqrt{2 + \omega})\,\, \cos r $ &
$[1 - 2 (2
+ \omega) \sin^2 r]\, \sec^2 r$ &
$ \sqrt{4+2\omega}\,\, \tan r$ & $(-2, \infty)$ \\
$(b\sqrt{-2}/\sqrt{2+\omega}) \,\,\cosh r$ &
$[1 + 2 (2
+ \omega) \sinh^2 r]\,{\rm sech}^2 r$
& $ \sqrt{2|\omega |-4}\,\, \tanh r $ & $ (-\infty, -2)$.
\end{tabular}
\end{table}

\begin{table}
\caption{The table shows the three different expressions for $\lambda (r)$ in
correspondence with those given in Table I. Their behavior
with the radial coordinate is also indicated.}
\begin{tabular}{cc}
$\lambda (r)$ & Typical feature of the solution \\ \hline
$(1+\omega) \ln[2(2+\omega)-b^2r^2] + \Upsilon r + {\cal C}$ &
Singularity at a finite proper distance.\\
$(1 + \omega) \ln(\cos r) + (2 +\omega) \,r^2/4$ & Naked
singularities at $r=n\pi/2$. \\
$(1 + \omega) \ln(\cosh r) - (2 +\omega)\,r^2/4$ & Well behaved
throughout the spacetime.
\end{tabular}
\end{table}

\clearpage

\begin{figure}[htbp]
\begin{center}
\epsfig{file=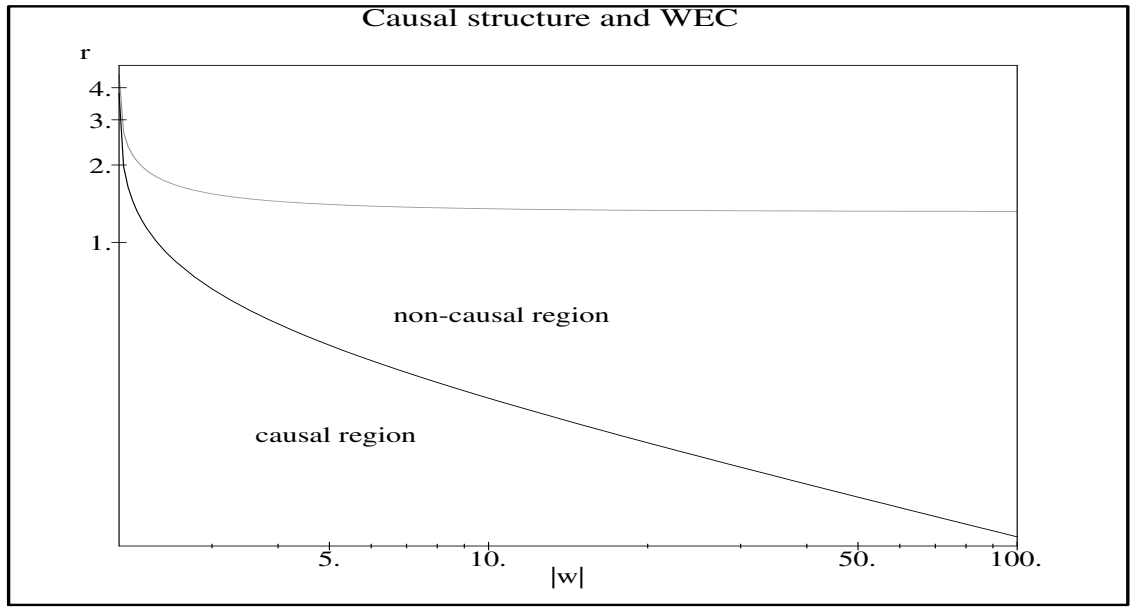,width=15cm,clip=}
\end{center}
\vspace{0.1cm}
\caption{Variation of $r_{\rm crit}$(black line) and $r_{\rm WEC}$ (grey line) with
$|\omega|$.}
\end{figure}

\end{document}